\documentclass[nofootinbib,superscriptaddress,longbibliography,aps,prd,reprint,amsmath,amssymb,preprintnumbers]{revtex4-2}
\usepackage{bm}
\usepackage{graphicx}
\usepackage{hyperref}
\usepackage{color}
\usepackage{siunitx}

\newcommand{\bfr}{\mathbf{r}}
\newcommand{\bfp}{\mathbf{p}}
\newcommand{\bfv}{\mathbf{v}}
\newcommand{\bfK}{\mathbf{K}}

\newcommand{\rmi}{\mathrm{i}}
\newcommand{\rmd}{\mathrm{d}}

\newcommand{\GF}{G_\mathrm{F}}
\newcommand{\weff}{\omega_\mathrm{eff}}
\newcommand{\wP}{\omega_\mathrm{P}}
\newcommand{\aN}{\alpha_\mathrm{N}}
\newcommand{\aC}{\alpha_\mathrm{C}}

\newcommand{\avg}[1]{\langle #1 \rangle}
\newcommand*{\figscale}{1}

\begin{document}

\title{Collisional flavor instability in dense neutrino gases}

\newcommand*{\UNM}{Department of Physics \& Astronomy, University of New Mexico, Albuquerque, New Mexico 87131, USA}

\author{Zewei Xiong}
\email{z.xiong@gsi.de}
\affiliation{GSI Helmholtzzentrum {f\"ur} Schwerionenforschung, Planckstra{\ss}e 1, 64291 Darmstadt, Germany}

\author{Lucas Johns}
\email{ljohns@berkeley.edu}
\affiliation{Departments of Astronomy and Physics, University of California, Berkeley, California 94720, USA}

\author{Meng-Ru Wu}
\email{mwu@gate.sinica.edu.tw}
\affiliation{Institute of Physics, Academia Sinica, Taipei 11529, Taiwan}
\affiliation{Institute of Astronomy and Astrophysics, Academia Sinica, Taipei 10617, Taiwan}
\affiliation{Physics Division, National Center for Theoretical Sciences, Taipei 10617, Taiwan}

\author{Huaiyu Duan}
\email{duan@unm.edu}
\affiliation{Department of Physics \& Astronomy, University of New Mexico, Albuquerque, New Mexico 87131, USA}

\begin{abstract}
    Charged-current neutrino processes such as $\nu_e + n \rightleftharpoons p + e^-$ and $\bar\nu_e + p \rightleftharpoons n + e^+$ destroy the flavor coherence among the weak-interaction states of a single neutrino and thus damp its flavor oscillation. In a dense neutrino gas such as that inside a core-collapse supernova or the black hole accretion disk formed in a compact binary merger, however, these ``collision'' processes can trigger large flavor conversion in cooperation with the strong neutrino-neutrino refraction. We show that there exist two types of collisional flavor instability in a homogeneous and isotropic neutrino gas which are identified by the dependence of their real frequencies on the neutrino density $n_\nu$. The instability transitions from one type to the other and exhibits a resonance-like behavior in the region where the net electron lepton number of the neutrino gas is negligible. In the transition region, the flavor instability grows exponentially at a rate $\propto n_\nu^{1/2}$. We find that the neutrino gas in the black hole accretion disk is susceptible to the collision-induced flavor conversion where the neutrino densities are the highest. Further investigations are needed to confirm if the collisional flavor instability will indeed result in the production of large amounts of heavy-lepton flavor neutrinos in this environment which would have important ramifications in its subsequent evolution.
\end{abstract}

\maketitle

\section{Introduction}
The proto-neutron star (PNS) in a core-collapse supernova (CCSN) and the black hole (BH) accretion disk/torus formed in a compact binary merger are immersed in and/or surrounded by dense neutrino media at birth. Emitted by the rapidly cooling remnants, these neutrinos help to shape the evolution of the physical environments through various processes including
\begin{align}
    \nu_e + n \rightleftharpoons p + e^-
    \quad\text{and}\quad
    \bar\nu_e + p \rightleftharpoons n + e^+.
    \label{eq:proc}
\end{align}
Because the neutrinos emitted by the remnants typically have different energy spectra and fluxes for different species, a change of the neutrino flavor, $\nu_e\rightleftharpoons\nu_{\mu/\tau}$ and $\bar\nu_e\rightleftharpoons\bar\nu_{\mu/\tau}$, can have important consequences in the remnant physics including but not limited to the nucleosynthesis in their ejecta. 

Although much has been learned in the last decade, the flavor evolution of the neutrinos in the compact object environment remains an unsolved problem. This is partly because, through the neutrino-neutrino refraction \cite{Fuller:1987aa,Notzold:1987ik,Pantaleone:1992xh}, a dense neutrino gas can experience flavor transformation collectively \cite{Samuel:1993uw,Pastor:2002we,Duan:2006jv,Duan:2010bg}. It is a daunting task to include both the neutrino flavor oscillations and the remnant dynamics in a single numerical simulation because of the large disparity between the time and distance scales of the two. But the integration of these two kinds of calculations is probably necessary given the discovery of the ``fast flavor conversion'' which can occur on the timescale of a nanosecond or shorter and even before the neutrinos are fully decoupled from the matter \cite{Sawyer:2015dsa,Tamborra:2020cul,Wu:2017qpc,Abbar:2018shq}. 

Recently, yet another type of collective flavor transformation is found that can take place inside the PNS or the BH accretion disk \cite{Johns:2021qby}. This kind of flavor transformation in a dense neutrino gas is triggered by the charged-current processes such as those in Eq.~\eqref{eq:proc}. These processes, when acting on a single neutrino, ``measure'' its flavor and, therefore, destroy the coherence among the different weak-interaction states of the neutrino. In a dense neutrino gas, however, these flavor decohering ``collisions'' can induce significant flavor conversion because the flavor evolution of different neutrino modes are tightly coupled by the neutrino-neutrino refraction.

\begin{figure}[htb!]
    \includegraphics[trim=1 2 1 1, clip, scale=\figscale]{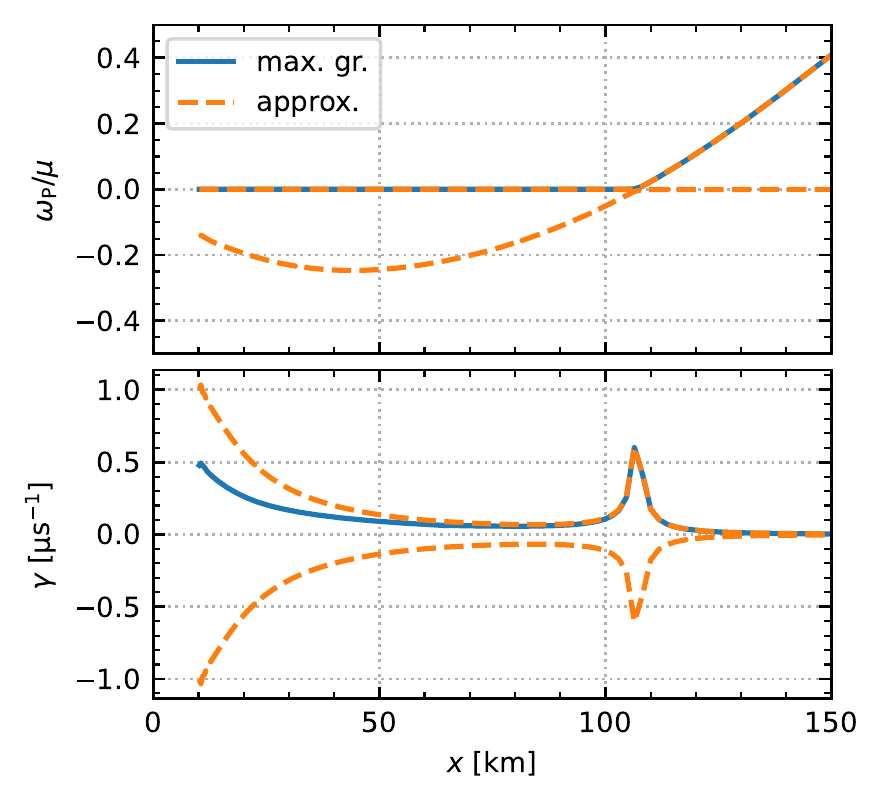}
    \caption{The frequency $\Omega=\wP+\rmi\gamma$ of the neutrino oscillation mode with the maximum growth rate (solid curves) in the equatorial plane of the BH accretion disk/torus model M3A8m3a5 at $t=\SI{20}{ms}$ \cite{Just:2014fka}. Also plotted is the analytic approximation of the frequencies of the two collective modes in the continuous energy limit (dashed curves) [Eq.~\eqref{eq:Omega-approx}]. The CFI of the plus type (with $\wP/\mu\approx0$) existing in the inner torus transitions to the minus type [with $\wP\propto (n_{\nu_e}-n_{\bar\nu_e})$] in the outer torus at $x\approx\SI{103}{km}$ where $n_{\nu_e}\approx n_{\bar\nu_e}$ and exhibits a resonance-like behavior with $\gamma\propto\mu^{1/2}$, where $\mu\propto n_{\nu_e}$.}
    \label{fig:20ms}
\end{figure}

The study of the collision-induced flavor conversion in Ref.~\cite{Johns:2021qby} was carried out for a monochromatic neutrino gas. The recent numerical simulation of the neutrino transport in a spherically symmetric supernova model (with tempered parameters) suggests that such flavor conversion can indeed occur inside the CCSN \cite{Xiong:2022vsy}. Very recently, the collisional flavor instability (CFI) has been shown to exist in the neutrino gas with a continuous energy spectrum  \cite{Lin:2022dek}. In this work, we further demonstrate that both types of CFI that were discovered in Ref.~\cite{Johns:2021qby} can exist in a homogeneous and isotropic neutrino gas and are identified by the frequencies of the corresponding neutrino oscillation modes. We will also show that the CFI of one type can transition to the other, e.g.\ in the hot BH accretion disk (Fig.~\ref{fig:20ms}), where the number densities of $\nu_e$ and $\bar\nu_e$ are similar.

\section{Equation of motion}
For simplicity, we consider the mixing of two neutrino flavors, the $e$ and $x$ flavors with the latter being a linear combination of the physical $\mu$ and $\tau$ flavor neutrinos. When the neutrinos are in the nearly pure weak-interactions states, the mean-field density matrix of the neutrino flavor \cite{Sigl:1993ctk} can be written as
\begin{align}
    \rho \approx \frac{f_{\nu_e} + f_{\nu_x}}{2} +
    \frac{f_{\nu_e} - f_{\nu_x}}{2}
    \begin{bmatrix}
        1 & S \\ S^* & -1
    \end{bmatrix}, 
\end{align}
where $f_{\nu_e}$ and $f_{\nu_x}$ are the occupation numbers in the corresponding weak-interactions states, respectively, and $S_{E,\bfv}(t,\bfr)$ (with $|S|\ll1$) is the flavor coherence of the neutrino of energy $E$ and velocity $\bfv$ at time $t$ and position $\bfr$. We assume that the neutrinos are ultra-relativistic so that $|\bfv|=c$, and we adopt the natural units with $\hbar=c=1$ throughout this work.   

The equation of motion at the linear order of $S$ is \cite{Izaguirre:2016gsx,Dasgupta:2021gfs}
\begin{multline}
    0= [\rmi v_\beta\partial^\beta - \sqrt2\GF v_\beta(j_e^\beta + j_\nu^\beta)+\weff] S_{E,\bfv}
    \\
    + \sqrt2\GF v^\beta\int_{-\infty}^\infty\frac{E'^2\rmd E'}{2\pi^2}\int\frac{\rmd\bfv'}{4\pi} v'_\beta G(E',\bfv') S_{E',\bfv'}.
    \label{eq:eom}
\end{multline}
In the above equation, $\GF$ is the Fermi constant, $j_e^\beta = \int\!(f_{e-} - f_{e^+}) u^\beta \rmd^3p/(2\pi)^3 $ is the four-current density of the net electron lepton number (ELN)  (in the absence of the heavy leptons), and $j_\nu^\beta = \int\![(f_{\nu_e} - f_{\nu_x}) - (f_{\bar\nu_e} - f_{\bar\nu_x})] v^\beta \rmd^3p/(2\pi)^3 $ is the $\nu$ELN four-current density, where $v^\beta = (1, \bfv)$ and $u^\beta$ are the four velocities of the neutrino and the charged lepton, respectively. In Eq.~\eqref{eq:eom}, we define the $\nu$ELN distribution as
\begin{align}
    G(E,\bfv) = \begin{cases}
        f_{\nu_e} - f_{\nu_x} & \text{ if } E > 0, \\
        f_{\bar\nu_x} - f_{\bar\nu_e} & \text{ if } E < 0,
    \end{cases}
\end{align}
where we denote the antineutrino as the neutrino with negative energy. We have also incorporated the effect of the flavor-decohering collisions \cite{Raffelt:1992uj} into the effective oscillation frequency 
\begin{align}
    \weff =  \cos(2\theta)\left(\frac{\Delta m^2}{2 E}\right) + \rmi\Gamma_E,
\end{align}
where $\Delta m^2$ and $\theta$ are the mass-squared difference and the vacuum mixing angle of the neutrino, respectively, and $\Gamma_E$ is the average of the emission and absorption rates of the electron flavor (anti)neutrino in Eq.~\eqref{eq:proc}. 

\section{Collisional flavor instability}
A collective flavor oscillation mode of the neutrino gas is given by $S_{E,\bfv}^{\Omega,\bfK}(t,\bfr) \propto e^{-\rmi(\Omega t - \bfK\cdot\bfr)}$ \cite{Duan:2008fd,Izaguirre:2016gsx}. In general, the frequency $\Omega=\wP + \rmi\gamma$ of a collective mode with the (real) wave vector $\bfK$ can be complex. A flavor instability is identified when $\gamma>0$ for a collective mode whose amplitude growths as $e^{\gamma t}$ until $|S|\sim 1$ or the physical conditions have changed significantly. In this work, we consider a homogeneous and isotropic gas which can exist in the compact object remnant where the neutrinos are trapped. We also focus on the collective mode that preserves the homogeneity and the isotropy, i.e.\ $S_{E,\bfv}^{\Omega,\bfK} \rightarrow S_{E}^{\Omega}$. Equation~\eqref{eq:eom} is greatly simplified in this case:
\begin{align}
    (\Omega + \weff) S_E^\Omega + \mu\int_{-\infty}^\infty g(E') S_{E'}^\Omega\,\rmd E'=0,
    \label{eq:coll}
\end{align}
where $\mu=\sqrt2\GF (n_{\nu_e}-n_{\nu_x})$ is a measure of the strength of the neutrino-neutrino refraction, and 
\begin{align}
    g(E) = \frac{E^2}{n_{\nu_e}-n_{\nu_x}}\int\!\frac{\rmd\bfv}{(2\pi)^3} G(E,\bfv)
\end{align}
is the energy $\nu$ELN distribution.
We have applied the shift $\Omega\rightarrow\Omega + \sqrt2\GF(j^0_e + j^0_\nu)$ in Eq.~\eqref{eq:coll} as is commonly done in the literature \cite{Izaguirre:2016gsx}. 

Equation~\eqref{eq:coll} implies $S_{E}^\Omega \propto 1/(\Omega + \weff)$ \cite{Banerjee:2011fj}, which can be substituted back into Eq.~\eqref{eq:coll} to obtain the self-consistency equation 
\begin{align}
    \int_{-\infty}^\infty\frac{g(E)\,\rmd E}{\Omega + \weff} = -\frac{1}{\mu}.
    \label{eq:Omega}
\end{align}
A solution to the above equation has been found in the large $\mu$ limit by assuming $\wP\propto\mu$ \cite{Lin:2022dek}. However, there are two branches of the dispersion relation $\Omega(\bfK)$ in a dense neutrino gas that preserve the axial symmetry about the wave vector $\bfK$  in the limit $\weff=0$  \cite{Izaguirre:2016gsx}. Therefore, one expects Eq.~\eqref{eq:Omega} to have two solutions that preserves the homogeneity (i.e.\ $\bfK=\mathbf{0}$) and isotropy in the limit $|\Omega|\gg|\weff|$. In this limit, we expand the left-hand side of Eq.~\eqref{eq:Omega} up to the first order of $\weff/\Omega$ and obtain
\begin{align}
    \Omega^2 + \mu  D  \Omega - \mu \avg{\weff} \approx 0,
    \label{eq:quadratic}
\end{align}
where 
\begin{align}
     D =\int_{-\infty}^\infty g(E)\,\rmd E = 1 - \frac{n_{\bar\nu_e}-n_{\bar\nu_x}}{n_{\nu_e}-n_{\nu_x}}
\end{align}
is a dimensionless measure of the net $\nu$ELN,
and $\avg{h}=\int_{-\infty}^\infty h(E)g(E)\,\rmd E$ is the average of an arbitrary function $h(E)$ weighted by the $\nu$ELN energy distribution. 

Equation~\eqref{eq:quadratic} indeed has two solutions
\begin{align}
    \Omega \approx \frac{-\mu D }{2} \pm \frac{\mu}{2}\sqrt{ D ^2 + 4\frac{\avg{\weff}}{\mu}}.
    \label{eq:Omega-approx}
\end{align}
In the limit $D^2 \ll |\avg{\weff}/\mu|$, we obtain
\begin{align}
    \Omega \approx \pm \sqrt{\mu\avg{\weff}}.
    \label{eq:Omega-res}
\end{align}
The above equation implies the existence of a flavor instability that grows at a rate $\gamma\propto \mu^{1/2}$ where the $\nu$ELN is negligible as long as $\avg{\Gamma_E}\neq0$.%
\footnote{An acute reader may realize the existence of a flavor instability in the collisionless gas if $\avg{\omega}<0$. This corresponds to the bipolar type of the neutrino oscillations \cite{Samuel:1993uw,Duan:2005cp} which can be understood as a flavor pendulum \cite{Hannestad:2006nj}.} 
In the opposite limit $ D^2\gg |\avg{\weff}/\mu|$, we find
\begin{align}
    \Omega_+ \approx \frac{\avg{\weff}}{D }
    \quad\text{and}\quad
    \Omega_- \approx -\mu D  - \frac{\avg{\weff}}{D }.
    \label{eq:Omega-pm}
\end{align}
We note that $\Omega_-$ is exactly the solution found in Ref.~\cite{Lin:2022dek} under the same conditions. We also note that these two solutions correspond to the two types of CFI in the monochromatic gas \cite{Johns:2021qby,Padilla-Gay:2022wck} which transition to one another as $D$ changes sign.

\begin{figure}[htb]
    \includegraphics[trim=1 2 1 1, clip, scale=\figscale]{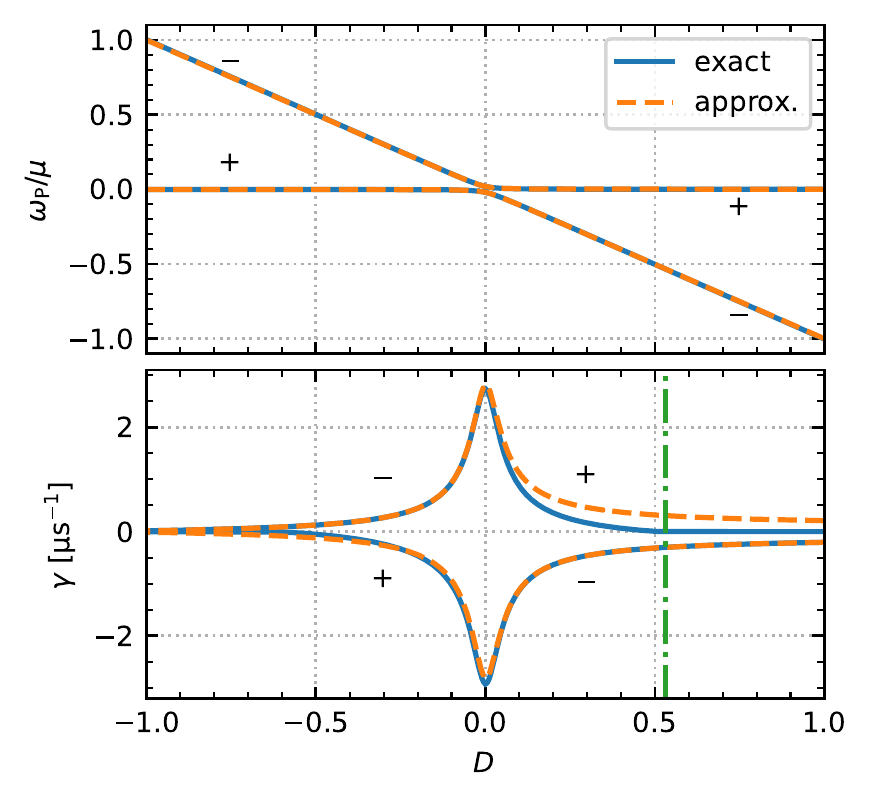}
    \caption{The collective frequencies $\Omega=\wP+\rmi\gamma$ in Eq.~\eqref{eq:Omega} (solid curves) and the approximate solutions in Eq.~\eqref{eq:Omega-approx} (dashed curves) with $\mu=\SI{e3}{\us^{-1}}$ and $\weff$ and $g(E)$ defined in Eqs.~\eqref{eq:weff-ex} and \eqref{eq:g-ex}, respectively. The solutions away from the resonance at $D=0$ are labeled as the plus and minus types according to Eq.~\eqref{eq:Omega-pm}. The vertical dot-dashed line in the lower panel corresponds to the approximate upper limit of $D\approx0.53$ of the instability which is obtained from Eq.~\eqref{eq:cond}  and is slightly higher than the actual upper limit $D\approx0.50$. The lower limit of the instability at $D\approx-1.13$ is not shown.}
    \label{fig:Omega}
\end{figure}

As a concrete example, we solve Eq.~\eqref{eq:Omega} with $\mu=\SI{e3}{\us^{-1}}$,
\begin{align}
    \weff = \SI{1.8}{\us^{-1}}\left(\frac{\SI{1}{MeV}}{E}\right)  
    + \frac{\rmi\aC}{\SI{e3}{\us}}\left(\frac{E}{\SI{1}{MeV}}\right)^2,
    \label{eq:weff-ex}
\end{align}
and
\begin{align}
    g(E) = \frac{2\aN}{3\zeta(3)} \left[\frac{E^2/T^3}{\exp(|E|/T)+1}\right],
    \label{eq:g-ex}
\end{align}
where $\aC=\aN=1$ and $T=\SI{4}{MeV}$ for the neutrino ($E>0$), and $\aC=0.3$, $\aN=D-1$, and $T=\SI{5}{MeV}$ for the antineutrino ($E<0$). We plot the real and imaginary parts of the frequencies of the two collective modes as functions of $D$ in Fig.~\ref{fig:Omega}. 
As comparison, we also plot the approximate solutions in Eq.~\eqref{eq:Omega-approx} as dashed curves in the same figure.

Figure~\ref*{fig:Omega} demonstrates a resonance-like instability where the net $\nu$ELN is negligible ($D^2\ll |\avg{\weff}/\mu|$) [Eq.~\eqref{eq:Omega-res}]. On the side of the resonance where $D<0$, the upper-branch solution is well approximated by $\Omega_-$. We shall call this flavor instability \textit{the minus type}, which has $\wP\propto\mu$. We call the instability on the other side of the resonance \textit{the plus type}, which has $\wP/\mu\approx 0$ in the large $\mu$ limit. Following Ref.~\cite{Lin:2022dek} we define
\begin{align}
    \Gamma = \frac{\int_0^\infty g(E) \Gamma_E \,\rmd E}{\int_0^\infty g(E)\,\rmd E}
    \quad\text{and}\quad
    \bar\Gamma = \frac{\int_{-\infty}^0 g(E) \Gamma_E \,\rmd E}{\int_{-\infty}^0 g(E)\,\rmd E}
\end{align}
as the average collision rates of the neutrino and the antineutrino, respectively. 
The numerical example shown in Fig.~\ref{fig:Omega} has $\Gamma/\bar\Gamma\approx2.13>1$, and the CFI exists in the range
\begin{align}
    \frac{\bar\Gamma}{\Gamma} \lesssim 1 - D = \frac{n_{\bar\nu_e}-n_{\bar\nu_x}}{n_{\nu_e}-n_{\nu_x}} \lesssim \frac{\Gamma}{\bar\Gamma}.
    \label{eq:cond}
\end{align}
The higher end of the inequality (or the lower limit of $D$) is determined by $\Omega_-$ in Eq.~\eqref{eq:Omega-pm}. We use the criterion in the monochromatic gas \cite{Johns:2021qby} for the lower end of the inequality (or the upper limit of $D$) because the condition $|\Omega|\gg|\weff|$ is no longer satisfied for the upper-branch solution at $D\gg|\avg{\weff/\mu}|$ in Fig.~\ref{fig:Omega}.
The direction of the inequality in Eq.~\eqref{eq:cond} is reversed if $\Gamma/\bar\Gamma<1$.

\begin{figure*}[htb!]
    \includegraphics[trim=1 2 1 1, clip, scale=\figscale]{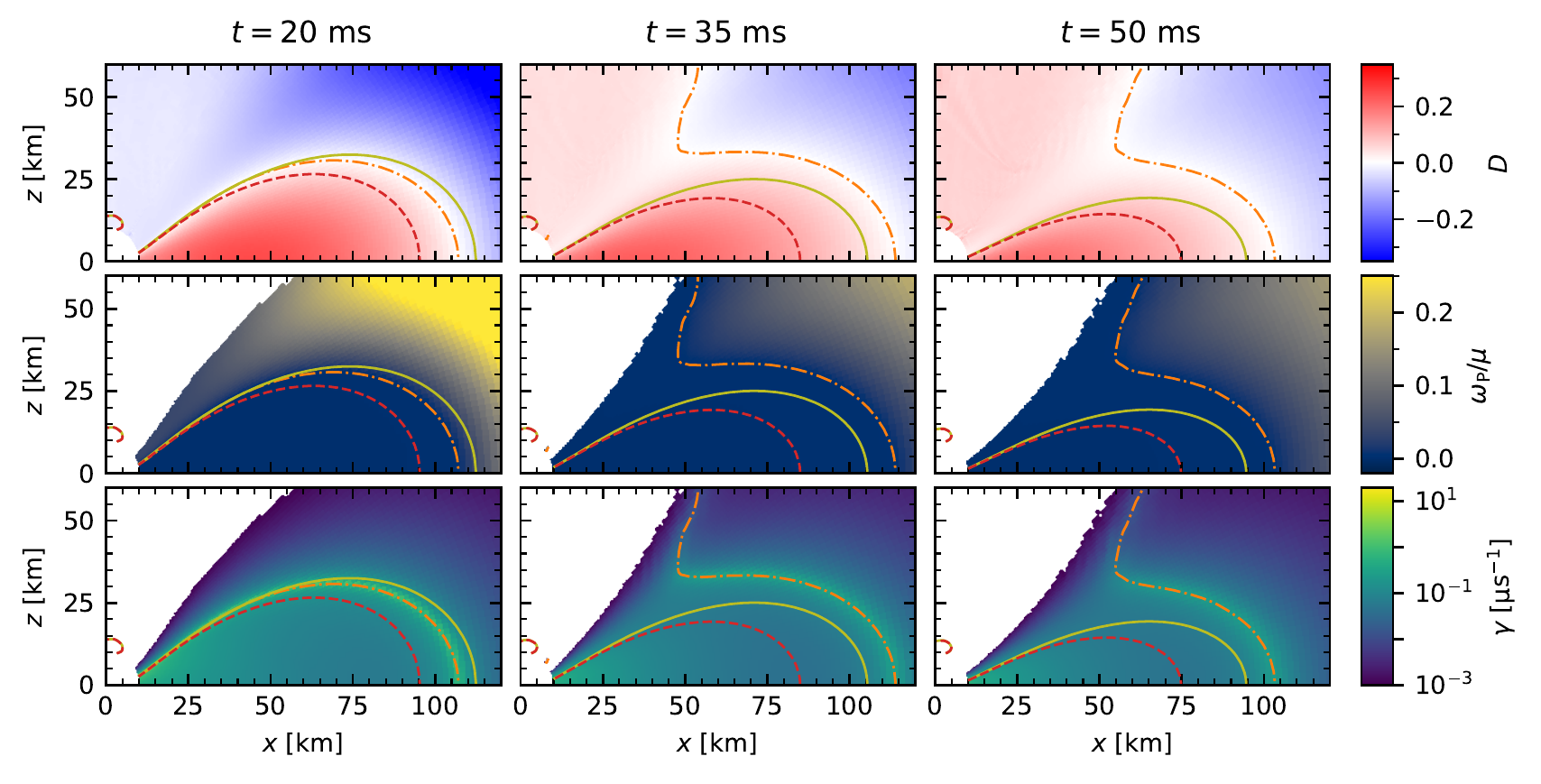}
    \caption{The $\nu$ELN excess parameter $D=1-n_{\bar\nu_e}/n_{\nu_e}$ (upper row) and the frequency $\Omega=\wP+\rmi\gamma$ of the normal mode with the maximum growth rate $\gamma>\SI{e-3}{\us^{-1}}$ (middle and lower rows) at three snapshots (as labeled) in the BH accretion disk model M3A8m3a5 of Ref.~\cite{Just:2014fka}. The solid and dashed curves are the contours with $F_{\nu_e}=1/3$ and $F_{\bar\nu_e}=1/3$, respectively. [See Eq.~\eqref{eq:F}]. The dot-dashed curves are the contours with $D=0$.}
    \label{fig:torus}
\end{figure*}

\section{Black hole accretion disk}
To investigate the existence of the CFI in the BH accretion disk/torus, we look into 
the model~M3A8m3a5 from \cite{Just:2014fka} which is a long-term axisymmetric simulation of a remnant BH-torus system including both the energy-dependent neutrino transport and a viscosity parameter. 
In the upper row of Fig.~\ref{fig:torus} we plot the net $\nu$ELN parameter $D=1-n_{\bar\nu_e}/n_{\nu_e}$ in three representative snapshots at $t=20$, 35, and \SI{50}{ms}, respectively. 
Following Ref.~\cite{Wu:2017drk}, we also plot
\begin{align}
    F_\nu(t,\bfr) = \frac{\left|\int\! \bfv\, f_\nu(t,\bfr,\bfp) \,\rmd^3 p\right|}{\int\! f_\nu(t,\bfr,\bfp)\, \rmd^3 p} = \frac{1}{3}
    \label{eq:F}
\end{align}
for $\nu=\nu_e$ and $\bar\nu_e$ as the solid and dashed curves, respectively. The condition of homogeneity and isotropy, which is one of the assumptions of this work, is approximately satisfied in the inner part of the disk where $F_\nu$ is small.%
\footnote{However, a recent study \cite{Liu:2023pjw} which appeared while this work was under review suggests that the qualitative results discussed here may still be valid even in anistropic environments.}
M3A8m3a5 does not include heavy-lepton neutrinos which have much smaller densities than $\nu_e$ and $\bar\nu_e$ in the BH-torus system.  

The high electron degeneracy in the inner disk favors the production of $\nu_e$ over $\bar\nu_e$ which leads to a positive $\nu$ELN. Above the emission surface, however, $\bar\nu_e$ generally has a higher flux than $\nu_e$ as a result of the protonization of the disk which is extremely neutron-rich. The only exception is near the polar region at later times where the larger emission surface of $\nu_e$ causes a higher concentration of $\nu_e$ than $\bar\nu_e$ \cite{Wu:2017drk}.
More physics details of the model can be found in Refs.~\cite{Just:2014fka,Wu:2017drk}.

For the neutrino gases with discrete energy groups, Eq.~\eqref{eq:coll} becomes
\begin{align}
    \sum_j\{[\Omega_a+\weff(E_i)]\delta_{ij} + \mu g_j \Delta E_j\} S_j^a = 0,
    \label{eq:coll-disc}
\end{align}
where $E_j$, $g_j$, and $\Delta E_j$ are the energy, the $\nu$ELN weight, and the width of the neutrino in the $j$th energy group, respectively. (The antineutrinos are counted as the neutrinos with  negative energies.) There are $N$ normal modes for $N$ discrete neutrino energy groups, and $\Omega_a$ and $S^a$ ($a=1,\ldots,N$) are the eigenvalues and eigenvectors of the matrix with the elements $\Lambda_{ij}=-[\weff(E_i)\delta_{ij} + \mu g_j \Delta E_j]$, respectively. We solve the frequencies of the normal modes in M3A8m3a5 with $\Delta m^2=\SI{2.5e-3}{eV^2}$, $\theta=\ang{8.6}$, and the emission and absorption rates of $\nu_e$ and $\bar\nu_e$ in Eq.~\eqref{eq:proc} \cite{Bruenn:1985en}. In Fig.~\ref{fig:torus} we show both the real and imaginary components of the frequency of the normal mode that has the largest growth rate in each spatial grid. One can see that the growth rates of the flavor instabilities are the largest where the net $\nu$ELN is negligible which is expected from the previous analysis.

Throughout the BH-torus system, one has $\Gamma/\bar\Gamma>1$ because the collision rates are dominated by the neutrino absorption rates and there are more neutrons than protons in this region.
Although the entire accretion disk tends to emit more $\bar\nu_e$ than $\nu_e$, the density of $\nu_e$ in the inner torus is actually larger where the chemical potential of the electron is significant.
Therefore, we expect only the CFI of the plus type (with $\wP/\mu\approx 0$) can exist in the inner torus. Earlier in Fig.~\ref{fig:20ms} we have shown the frequency $\Omega$ of the normal mode with the largest growth rate in the equatorial plane of the accretion disk at $t=\SI{20}{ms}$. It is clear from Figs.~\ref{fig:20ms} and \ref{fig:torus} that the CFI in the inner torus is indeed of the plus type, while the instability in the outer region of the torus is of the minus
type [$\wP\propto(n_{\nu_e}-n_{\bar\nu_e})$] if our analysis can be generalized to the anisotropic environment. 

\section{Discussion and conclusions}
We have shown that there exist two types of CFI in a dense neutrino gas that preserves the homogeneity and isotropy. The CFI transitions from one type to the other where the net $\nu$ELN is zero and has a resonance-like instability that grows at a rate $\propto n_{\nu}^{1/2}$. But this is only part of the story. There can exist the CFI that breaks these symmetries or even in the inhomogeneous and anisotropic environment as one maps out the full dispersion relation $\Omega(\bfK)$ of the collective neutrino oscillation wave. Numerical simulations are needed to confirm if the existence of the CFI can lead to significant flavor conversion before the physical conditions change.

We would like to emphasize that, although the CFI can interplay with the fast neutrino flavor conversion \cite{Johns:2021qby, PhysRevD.106.103029, Padilla-Gay:2022wck}, its existence does not require a crossing of the $\nu$ELN angular distribution as the latter does. Therefore, the CFI can exist in CCSNe and compact binary merger remnants in the regions and at the epochs where/when fast flavor instabilities do not yet exist (see, e.g.\ Ref.~\cite{Just:2022flt}).

Our result is strong motivation for follow-up work assessing how the CFI impacts the dynamics, element production, and kilonova emission of neutron-star mergers. For example, substantially more $\nu_{\mu/\tau}$ and $\bar\nu_{\mu/\tau}$ can be produced by the BH accretion disk through flavor conversion and thus cool the remnant faster. The change of the physical conditions will, of course, affect the existence of the CFI as well as the condition of fast flavor instabilities. 
This again cries out for the integration of the flavor oscillations into the neutrino transport in the simulations of CCSNe and compact binary mergers.  

\begin{acknowledgments}
We thank O.~Just for providing the data of model M3A8m3a5 and N.~Raina for pointing out several typos in an earlier version of the manuscript. H.~D.\ is supported by the US DOE NP grant No.\ DE-SC0017803 at UNM.
M.-R.~W.\ acknowledges supports from the National Science and Technology Council, Taiwan under Grant No.~110-2112-M-001-050 and 111-2628-M-001-003-MY4, the Academia Sinica under Project No.~AS-CDA-109-M11, and Physics Division, National Center for Theoretical Sciences, Taiwan.
Z.~X.\ acknowledges support of the European Research Council (ERC) under the European Union’s Horizon 2020 research and innovation programme (ERC Advanced Grant KILONOVA No. 885281).
L.~J.\ acknowledges support from NASA Hubble Fellowship grant number HST-HF2-51461.001-A awarded by the Space Telescope Science Institute, which is operated by the Association of Universities for Research in Astronomy, Incorporated, under NASA contract NAS5-26555.
\end{acknowledgments}

\bibliography{refs}

\end{document}